\documentstyle[12pt]{article}

\begin{document}

{}
\hfill   {\bf \large IFT/32/99}
\vskip1in
\centerline{\Large {\bf Running Couplings in Hamiltonians$^\dagger$}}
\vskip .1in
\centerline{\small {December, 1999}}
\vskip .3in
\centerline{Stanis{\l}aw D. G{\l}azek}
\vskip .1in
\centerline{Institute of Theoretical Physics, Warsaw University}
\centerline{ul. Ho{\.z}a 69, 00-681 Warsaw}
\vskip.5in

\centerline{\bf Abstract}
\vskip.1in

We describe key elements of the perturbative similarity renormalization
group procedure for Hamiltonians using two, third-order examples:
$\phi^3$ interaction term in the Hamiltonian of scalar field theory in 6
dimensions and triple-gluon vertex counterterm in the Hamiltonian of QCD
in 4 dimensions.  These examples provide insight into asymptotic freedom
in Hamiltonian approach to quantum field theory.  The renormalization
group procedure also suggests how one may obtain ultraviolet-finite effective
Schr\"odinger equations that correspond to the asymptotically free
theories, including transition from quark and gluon to hadronic degrees
of freedom in case of strong interactions.  The dynamics is invariant
under boosts and allows simultaneous analysis of bound state structure
in the rest and infinite momentum frames.

\vskip.5in
\vskip2.2in

$^\dagger$ {\it This article grew out of an invited talk at The Workshop
on Light-Cone QCD and Nonperturbative Hadron Physics, Centre for The
Subatomic Structure of Matter and The National Institute of Theoretical
Physics, University of Adelaide, Adelaide, Australia, December 13-23,
1999.}

\newpage

{\bf 1. INTRODUCTION}
\vskip.1in

Canonical Hamiltonians of quantum field theories can be written in
terms of creation and annihilation operators. Let us denote those
operators by $q$. In this notation, for example, a triple-gluon
vertex in the light-front QCD Hamiltonian has the structure

$$ H_Y \quad = \quad \sum_{123} \int
[123] \,\delta(1 + 2 - 3) \,g \,Y_{123}\,\,q^\dagger_1\,
q^\dagger_2\, q_3 \quad + \quad h.c. \quad ,  \eqno(1.1) $$

\noindent where symbols $1$, $2$ and $3$ denote colors, spins or momenta
of gluons, $[123]$ is a shorthand notation for the integration measure
over the gluon three-momenta, $\delta(1 + 2 - 3)$ is the Dirac
$\delta$-function of three-momentum conservation, and $Y_{123}$ is a
function of the gluon quantum numbers, implied by QCD.  $g$ denotes a
bare canonical coupling constant, which is expected from Lagrangian
approach to require ultraviolet renormalization.  However, the
Hamiltonian lacks many of the Lagrangian density symmetries that are
employed in the perturbative Lagrangian renormalization procedure and, in all
light-front Hamiltonians, transverse and longitudinal directions are
treated differently.  For that reason, and to introduce a method for
renormalization of Hamiltonians such as (1.1), we need to review the
origin of the ultraviolet divergences.

One can evaluate matrix elements of $H_Y$ between states of the form
$|ijk...\rangle = q^\dagger_iq^\dagger_jq^\dagger_k . . . |0\rangle$.
Consider $\langle 12|H_Y|3\rangle$, which is proportional to $Y_{123}$.
The trouble is that $Y_{123}$ does not vanish when the relative motion
of particles 1 and 2 becomes very energetic.  In other words, the
interaction Hamiltonian directly couples states of small kinetic energy
to states of arbitrarily high kinetic energy.  For example, when we
square $H_Y$ and evaluate $\sum_{i}H_Y |i\rangle \langle i|H_Y$, the
range of $|i\rangle$s to sum over, on the energy scale, is infinite.
The sum diverges, and the square of $H_Y$ does not exist.  If one tries
to find eigenstates of the Hamiltonian, the divergence will dominate
finite terms.  Also, $\exp{-iHt}$ diverges and no conclusions can be
drawn from knowing $H$ as it stands.  The problem is worse than the
inifinite energy range of interaction implies by itself:  the function $Y_{123}$
in QCD grows when the energy difference between kinetic energies of
particles 1 and 2, and particle 3 grows.  The larger are the kinetic
energies of the intermediate particles, the more important become the
interactions, and we are sent into an abyss of Fock space states without
bounds.  All physically relevant local quantum field theories have this
trouble.

To see the essence of the problem, imagine a matrix of the Hamiltonian
matrix elements in the basis of eigenstates of certain $H_0$,
$H_0|i\rangle = E_{0i}|i\rangle$, $H_{ij}\equiv \langle i|H|j\rangle$.
Our problem is that the

 \vskip.1in
\begin{figure}[h]
\begin{picture}(400,120)(-10,0)
\thicklines
\put(190,0){\line(1,0){120}}
\put(190,0){\line(0,1){120}}
\put(310,120){\line(-1,0){120}}
\put(310,120){\line(0,-1){120}}
{\large
\put(10,55){$H_{ij} \quad = \quad H_{0 ij}+ H_{Y ij} \quad = \quad $}
}
\put(320,0){$low~E_{0 i}$}
\put(320,115){$high~E_{0 i}$}
\put(190,-15){$high~E_{0 j}$}
\put(280,-15){$low~E_{0 j}$}
\put(265,110){high-low}
\put(195,110){high-high}
\put(195,5){low-high}
\put(270,5){low-low}
\end{picture}
\end{figure}
\vskip.05in

\noindent corners marked $low-high$ and $high-low$ on the above figure,
contain too large matrix elements for the Hamiltonian matrix to have
eigenvalues that are independent of the matrix boundaries.  In order to
understand the boundary-dependence problem of the eigenvalues, we put an
upper bound, denoted by $\Delta$, on the basis states energies and we
work out what happens with eigenstates of Hamiltonian matrices whose
size is limited by the conditions $E_{0i} \le \Delta$ and $E_{0j} \le
\Delta$, when we change $\Delta$.  In particular, we ask what properties
must $H_Y$ have for the spectrum of $H_\Delta$ to have a limit when
$\Delta \rightarrow \infty$.

Wilson asked this question in case of a model Hamiltonian with big
energy gaps between successive energy scales that were included in his calculation, and he studied the influence of coupling between small and large energy states on the lowest eigenvalues. \cite{Wilson1} His method for dealing with the $\Delta$-dependence of the spectrum in the multiscale eigenvalue problem (in general, the problem of dependence on regularization, of any kind),
was based on an iterative procedure.  Initially, one solves the highest
energy part of dynamics and focuses on its lowest eigenvalue levels whose dynamics, in turn, is dominated by states with energies lower by one energy gap.  Then, one solves this next lower energy scale dynamical problem, and one repeats the process many times. Starting from the energy scale $\Delta$, one eventually arrives at a finite scale $\lambda$.  This is schematically indicated on the following figure.

\vskip.2in
\centerline{\bf Wilson's approach }
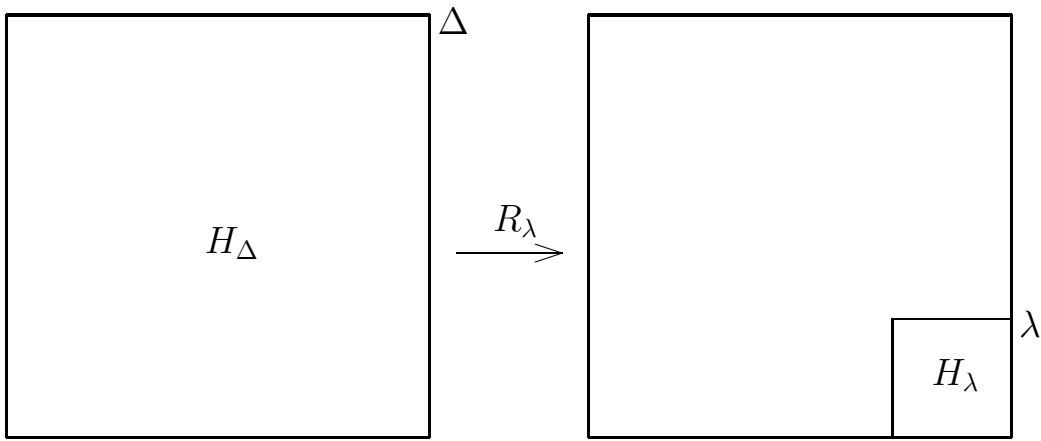
\begin{figure}[h]
\begin{center}
{\large
\begin{picture}(400,170)(-10,0)

\thicklines
\put(-10,0){\line(1,0){160}}
\put(-10,0){\line(0,1){160}}
\put(150,160){\line(-1,0){160}}
\put(150,160){\line(0,-1){160}}

\put(210,0){\line(1,0){160}}
\put(210,0){\line(0,1){160}}
\put(370,160){\line(-1,0){160}}
\put(370,160){\line(0,-1){160}}

\thinlines
\put(325,45){\line(1,0){45}}
\put(325,45){\line(0,-1){45}}
\put(153,153){$\Delta$}
\put(373,38){$\lambda$}

\put(160,70){\line(1,0){40}}
\put(200,70){\line(-3,1){10}}
\put(200,70){\line(-3,-1){10}}
\put(174,78){$R_\lambda$}

\put(65,70){$H_\Delta$}
\put(340,20){$H_\lambda$}

\end{picture}
}
\end{center}
\end{figure}

In this figure, we see a new small matrix of size $\lambda$, denoted by
$H_\lambda$.  This matrix is calculated using an operation $R_\lambda$,
which is constructed in the sequence of steps lowering the cutoff from
$\Delta$ to $\lambda$.  The construction is based on the principle that
the smallest eigenvalues of the small matrix, $H_\lambda$, should be the
same as the smallest eigenvalues of the big matrix $H_\Delta$.  The
algebraic derivation of the small Hamiltonian $H_\lambda$ is designed to
guarantee the equality of the smallest eigenvalues.  The crux is that if
all matrix elements of $H_\lambda$ are independent of $\Delta$ (in
general, independent of the regularization one uses to define
$H_\Delta$) then, the eigenvalues of $H_\lambda$ must be independent of
the regularization.  Therefore, if we know what to do with the
regularization dependence of matrix elements of $H_\lambda$, then we
know how to go about regularization dependence of the spectrum of
$H_\Delta$.

Note, that $\lambda$ is finite and can be chosen arbitrarily, as long as
$\Delta \rightarrow \infty$.  The set of transformations that connect
Hamiltonians with different values of $\lambda$ is called
renormalization group.  \cite{Wilson2} Physical results should be
independent of $\lambda$, by construction.  It is clear that $H_\lambda$
cannot be equal to merely $P_\lambda H_\Delta P_\lambda$, where
$P_\lambda$ denotes a projection operator that projects on the space of
states with energies $E_{0i} < \lambda$.  Some additional terms must be
included, which reproduce dynamical effects from above $\lambda$.
Similarly, we do not expect that $H_\Delta$ is merely $P_\Delta
H_{canonical} P_\Delta$, where $H_{canonical}$ is built from $H_0$ and
terms such as (1.1).  By the same argument as for $H_\lambda$, some
additional terms are required in $H_\Delta$, to include effects from
above the cutoff $\Delta$.  We will call those terms {\it counterterms}.
The problem is how to find them.  According to Wilson, they are found
from the condition that all matrix elements of $H_\lambda$ become
independent of the regularization in the limit $\Delta \rightarrow
\infty$, for all finite values of $\lambda$.  One must take care not
only of the diverging (i.e.  $\Delta$-dependent) regularization
dependence, but also of the finite regularization effects. The coupling constant $g$ in Eq. (1.1) is changed in the renormalization process as a result of introducing counterterms.

The problem with the transformation $R_\lambda$ is that the large energy
gaps are absent in physically relevant cases, and perturbation theory
based on energy scales alone fails.  One can try to take advantage of a
small coupling constant but we know that naive perturbative expansion
does not work in degenerate cases.  For example, one may think of
effects familiar from elementary degenerate perturbation theory for
eigenvalues of a Hamiltonian matrix with a few rows and columns.  We
know that perturbation theory cannot work unless one properly chooses
the initial basis states in the degenerate subspace - eigenstates of the interaction matrix (if it is a few rows
and columns).  Only in
that basis the perturbative limit $g \rightarrow 0$ exists.  Otherwise
perturbation theory produces vanishing energy denominators that lead to
diverging terms and the calculation is misleading.  In case of quantum
field theories of interest, the situation is much more involved than in simple matrix case due to multiply degenerated continuous spectra of $H_0$.  Moreover, in asymptotically free theories, we expect that the interaction strength
grows when we go from $H_\Delta$ to $H_\lambda$ and small energy
denominators are certainly expected to produce large effects.  Thus, the
degeneracy of spectra and strength of couplings do not allow us to do a
precise analysis of $R_\lambda$ and $H_\lambda$.  So, the operation
$R_\lambda$ is of limited applicability in the canonical approach to
quantum field theory.

\vskip.3in
{\bf 2. SIMILARITY FOR HAMILTONIANS}
\vskip.1in

There is an alternative approach \cite{GlazekWilson1}
\cite{GlazekWilson2} (see the figure below).  Instead of calculating a
small Hamiltonian matrix, we can also calculate a narrow matrix.
Namely, a similar (in the sense of algebraic similarity) matrix that has
the same eigenvalues but whose matrix elements $H_{\lambda i j}$ vanish
if $|E_{0i} - E_{0j}| > \lambda$ (or another condition of ``narrowness''
is satisfied - hermitian matrices can be diagonalized and, therefore,
partial diagonalization to a narrow matrix should be possible).  The
choice of near-diagonal form is motivated by the following property of
near-diagonal matrices:  when we act with them on a state of some finite
energy, a single action of the matrix can rise the energy by at most
$\lambda$, i.e. by its width on the energy scale.  In perturbation
theory for eigenvalues of a near-diagonal matrix, corrections will not
be sensitive to the cutoff $\Delta$ up to the order $n \sim
\Delta/(2\lambda)$, since one has to go up in energies and come back to
the initial energy range through action of interactions, and at least
$n$ are needed to reach the boundary starting from $\lambda$.  In
similarity, the crux is that if matrix elements of the narrow matrix of
finite width $\lambda$ are independent of regularization when $\Delta
\rightarrow \infty$, then the spectrum of $H_\lambda$ will be
independent of regularization to all orders of perturbation theory.

\vskip.2in
\centerline{\bf Similarity approach}
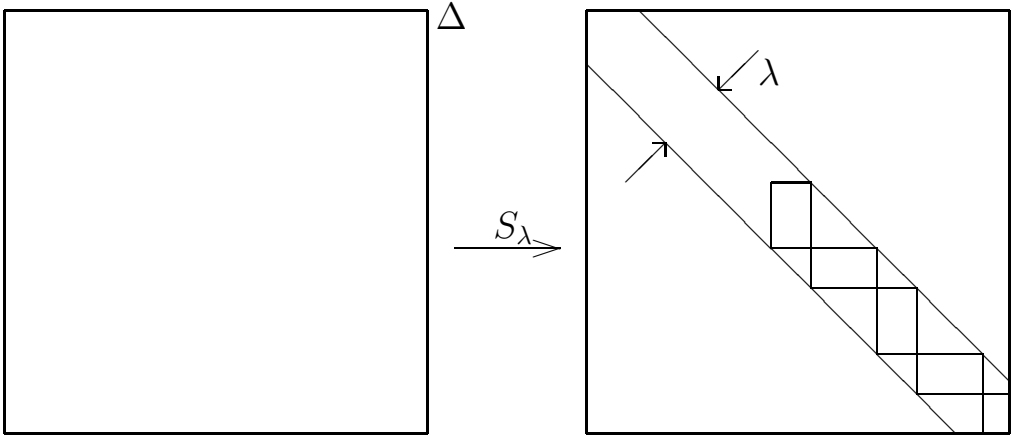
\begin{figure}[h]
\begin{center}
{\large
\begin{picture}(400,170)(-10,0)

\thicklines
\put(-10,0){\line(1,0){160}}
\put(-10,0){\line(0,1){160}}
\put(150,160){\line(-1,0){160}}
\put(150,160){\line(0,-1){160}}
\put(210,0){\line(1,0){160}}
\put(210,0){\line(0,1){160}}
\put(370,160){\line(-1,0){160}}
\put(370,160){\line(0,-1){160}}

\thinlines
\put(350,0){\line(-1,1){140}}
\put(370,20){\line(-1,1){140}}

\put(240,110){\line(-1,-1){15}}
\put(260,130){\line(1,1){15}}
\put(240,110){\line(-1,0){5}}
\put(240,110){\line(0,-1){5}}
\put(260,130){\line(1,0){5}}
\put(260,130){\line(0,1){5}}


\put(370,15){\line(-1,0){35}}
\put(335,15){\line(0,1){40}}
\put(335,55){\line(-1,0){40}}
\put(295,55){\line(0,1){40}}
\put(295,95){\line(-1,0){15}}
\put(280,95){\line(0,-1){25}}
\put(280,70){\line(1,0){40}}
\put(320,70){\line(0,-1){40}}
\put(320,30){\line(1,0){40}}
\put(360,30){\line(0,-1){30}}

\put(153,153){$\Delta$}
\put(275,132){$\lambda$}

\put(160,70){\line(1,0){40}}
\put(200,70){\line(-3,1){10}}
\put(200,70){\line(-3,-1){10}}
\put(175,74){$S_\lambda$}

\end{picture}
}
\end{center}
\end{figure}

The transformation $S_\lambda$ is called
similarity transformation.  The effective Hamiltonian matrices with
various widths $\lambda$ are connected by transformations that are
called similarity renormalization group transformations for
Hamiltonians.  We limit our discussion to unitary transformations
$S_\lambda$.  The key feature of the similarity approach is that
perturbative construction of $S_\lambda$ avoids small energy
denominators entirely - they are limited from below by the width
$\lambda$.  In turn, the perturbatively calculated narrow Hamiltonians
can be diagonalized numerically, which is the ultimate way to find
solutions to complex non-perturbative problems of the original theory.
One could ask, why don't we go all the way to $\lambda = 0$, which would
mean complete diagonalization through $S_\lambda$ with $\lambda=0$?
This is impossible in perturbation theory.  We can trust perturbation
theory for calculating $H_\lambda$ only for not too small values of
$\lambda$.

In other words, the perturbative similarity transformation $S_\lambda$
involves energy changes that are at least as large as $\lambda$ and the
problem with large effects in perturbative evaluation of effective
Hamiltonians is overcome.  But that does not mean we eliminated any of
the nonperturbative effects.  They are still hidden in the narrow
effective Hamiltonian, as much as they were in the initial one.  The
only thing we accomplish through similarity, is the elimination of
direct couplings between states of interest to us and very high energy
states.  This is a prerequisite that we need to define an ultraviolet
finite, nonperturbative Hamiltonian eigenvalue problem in quantum field
theory.

One can apply various perturbative procedures for calculating
$S_\lambda$ and $H_\lambda$. Wegner invented a beautifully simple scheme 
for evaluating near diagonal Hamiltonians in solid state physics. \cite{Wegner} It was shown \cite{Australia} that Wegner's equation can
be employed in the renormalization group scheme.  A number of variations
reported in the literature in different areas, is growing.  \cite{RofH}

The bottom line is that when solving for the spectrum of a narrow
Hamiltonian, we do not have to diagonalize the whole matrix of size
$\Delta \rightarrow \infty$.  We can select a window, as is illustrated
in the next figure.  Diagonalization of that small window is much
simpler than diagonalization of the whole matrix and the window
eigenvalues match the whole matrix eigenvalues in the middle range of
window energies.  \cite{GlazekWilson3} The reason is that the wave
functions have width comparable to $\lambda$ on the energy scale, which
is indicated on the figure (see also Appendix B in Ref.
\cite{GlazekWilson1}).  One cannot expect wave functions of eigenstates
of the initial Hamiltonian matrix, $H_\Delta$, to be dominated by some
small range of energies.  In contrast, the eigenstates of matrix
$H_\lambda$ are expected to have wave functions that have this property.
Recent calculations of quarkonium and glueball spectra in light-front
QCD exploit this feature.  \cite{BrisudovaPerryWilson}
\cite{AllenAdelaide}

\vskip.2in
\centerline{\bf Window Hamiltonians}
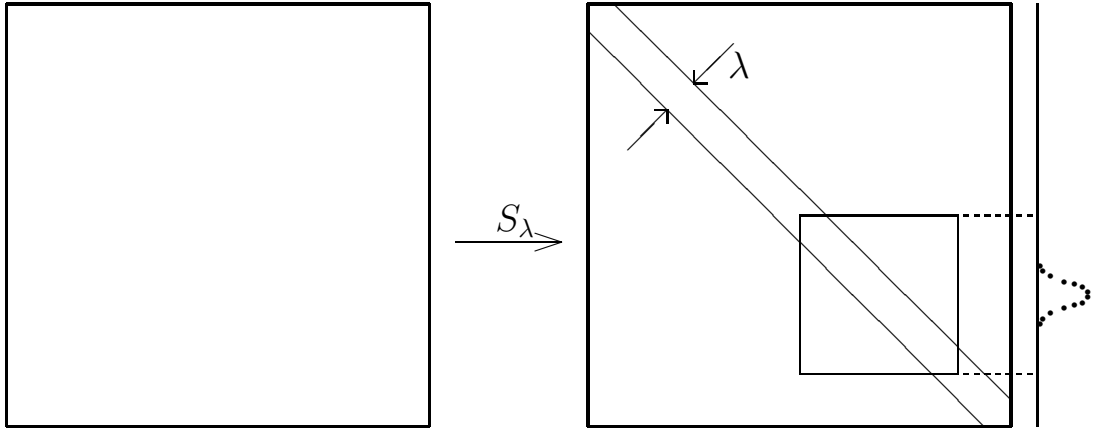
\begin{figure}[h]
\begin{center}
{\large
\begin{picture}(410,170)(-10,0)

\thicklines
\put(-10,0){\line(1,0){160}}
\put(-10,0){\line(0,1){160}}
\put(150,160){\line(-1,0){160}}
\put(150,160){\line(0,-1){160}}

\put(210,0){\line(1,0){160}}
\put(210,0){\line(0,1){160}}
\put(370,160){\line(-1,0){160}}
\put(370,160){\line(0,-1){160}}

\put(380,0){\line(0,1){160}}

\thinlines
\put(360,0){\line(-1,1){150}}
\put(370,10){\line(-1,1){150}}
\put(240,120){\line(-1,-1){15}}
\put(250,130){\line(1,1){15}}
\put(240,120){\line(-1,0){5}}
\put(240,120){\line(0,-1){5}}
\put(250,130){\line(1,0){5}}
\put(250,130){\line(0,1){5}}
\put(263,132){$\lambda$}

\put(160,70){\line(1,0){40}}
\put(200,70){\line(-3,1){10}}
\put(200,70){\line(-3,-1){10}}
\put(175,74){$S_\lambda$}

\put(290,20){\line(1,0){60}}
\put(290,20){\line(0,1){60}}
\put(350,80){\line(-1,0){60}}
\put(350,80){\line(0,-1){60}}

\multiput(352,80)(4,0){7}{\line(1,0){2}}
\multiput(352,20)(4,0){7}{\line(1,0){2}}

\put(381,61){\circle*{2}}
\put(382,59){\circle*{2}}
\put(385,57){\circle*{2}}
\put(390,55){\circle*{2}}
\put(394,54){\circle*{2}}
\put(397,53){\circle*{2}}
\put(399,51){\circle*{2}}
\put(399,49){\circle*{2}}
\put(397,47){\circle*{2}}
\put(394,46){\circle*{2}}
\put(390,45){\circle*{2}}
\put(385,43){\circle*{2}}
\put(382,41){\circle*{2}}
\put(381,39){\circle*{2}}


\end{picture}
}
\end{center}
\end{figure}

Ref.  \cite{GlazekWilson3} outlines the similarity procedure starting
from an initial Hamiltonian, through evaluation of the narrow
Hamiltonians (having found the necessary counterterms in the initial
Hamiltonian) in perturbation theory, to diagonalization of a small
window to get the bound state energy, using a matrix example.  The
matrix model is asymptotically free and has a bound state.  The coupling
constant is a function of the effective Hamiltonian width $\lambda$, we
say it ``runs''.  For example, it may equal about 0.06 at $\lambda = 65$
TeV and about 1 at 1 GeV.  Still, the window Hamiltonian of a few GeV size 
can be calculated in second order perturbation theory and the window bound
state eigenvalue deviates from exact solution by only 10\%.  Once we
understand that example, we can return to quantum field theories with
interactions of the form (1.1) and attempt a calculation of the
corresponding ``window'' Hamiltonians, with running couplings.
Calculations can be carried out using the notion of effective particles.

\vskip.3in
{\bf 3. SIMILARITY FOR PARTICLES}
\vskip.1in

When we deal with huge Hamiltonian matrices of quantum field theory the
number of states is as big as we let it be and the number of matrix
elements we have to think about becomes very quickly incomprehensible.
We have to reduce the amount of information that we need to know at the
beginning.  Imagine we would know matrix elements of the interaction
$\alpha/r$ in atomic basis functions, numerically, but we would not know
that they all correspond to the Coulomb force.  It would be very hard to
relate what happens in one atomic system to what happens in another one.
Therefore, when we aim at universal calculations of effective
Hamiltonians in theories that contain interactions such as (1.1), we may
proceed to a new version of similarity transformation, which avoids
dealing directly with Hamiltonian matrix elements in a particular basis
and, instead, operates at the level of field operators.
\cite{GlazekAPP}

Let us introduce a unitary transformation ${\cal U}_\lambda$ that
transforms field operators (denoted here by $\phi$, independently of
their spin or other quantum numbers they carry),

$$ \phi_\lambda (x) \quad = \quad {\cal U}_\lambda \,\, \phi_\infty (x) \,\,
{\cal U}^\dagger_\lambda \quad . \eqno(3.1) $$

\noindent $\phi_\infty (x)$ denotes a bare quantum field operator that,
at any prescribed time, can be expanded into creation and annihilation
operators for bare particles in a canonical fashion that we do not need
to define here very precisely.  $\phi_\lambda (x)$ denotes an operator
that is built in exactly the same way from creation and annihilation
operators for effective (dressed) particles.  This kind of
transformation is motivated by physics of hadrons, whose structure can
be explained in a constituent quark model.  Dressed particles in a given
theory interact differently than the bare ones.  Namely, bare ones have
interactions like (1.1), while the effective ones can only exchange
momentum transfers that are limited by $\lambda$.  This is secured by
the construction of ${\cal U}_\lambda$, to be explained below.
Therefore, {\it the effective particle wave functions of eigenstates of
the Hamiltonian may quickly fall off when momenta or number of the
effective particles deviate from the physically dominant values.} This is
why one can hope to obtain a constituent picture of hadrons in QCD using
similarity for particles. More generally, the expected convergence of eigenstate expansion in effective particle basis in Fock space opens a door to studies 
of few-body systems in quantum field theory.

In order to set up equations that will allow us to calculate
Hamiltonians for effective particles, let's rewrite Eq.  (3.1) in terms
of the creation and annihilation operators,

$$ q_\lambda \quad = \quad {\cal U}_\lambda \,\, q_\infty \,\, {\cal
U}^\dagger_\lambda \quad . \eqno(3.2) $$

\noindent All we need to do next is:  take the bare Hamiltonian of our
theory, as it is given initially in terms of $q_\infty$, calculate
counterterms it needs to contain in addition to the canonical terms,
obtain this way our initial $H_\infty(q_\infty)$, and rewrite it in
terms of $ q_\lambda$.  ${\cal U}_\lambda$ is secured to be unitary by
construction.  The whole point of the construction is that the resulting
$H_\lambda(q_\lambda)$ is to contain only such interaction terms that,
when we evaluate their matrix elements between Fock basis states of
effective particles, the resulting effective Hamiltonian matrix is
narrow, of width $\lambda$, as in the similarity procedure for
Hamiltonian matrices we discussed in previous Section.  It will not be
necessary to go into details here.  Only a brief outline of the scheme
follows.

Since rewriting the Hamiltonian in different degrees of freedom does not
change the operator itself, we have $H_\lambda(q_\lambda) =
H_\infty(q_\infty)$.  One may think about $H_\lambda(q_\lambda)$ as a QCD
Hamiltonian written in terms of constituent quarks and gluons, and
about $H_\infty(q_\infty)$ as the same QCD Hamiltonian written in terms of
canonical quarks and gluons, associated with partons, or current quarks
(to make the connection between hadronic rest frame constituents and partons in the infinite momentum frame, we have to use the light-front form of Hamiltonian dynamics, see \cite{Wilsonetal} for an outline of light-front QCD in the context of renormalization group procedure for Hamiltonians).

Applying the transformation ${\cal U}_\lambda$, one obtains ${\cal
H}_\lambda \equiv H_\lambda(q_\infty) = {\cal U}^\dagger_\lambda
H_\infty(q_\infty) {\cal U}_\lambda$.  This relation means that the
operator ${\cal H}_\lambda$ has the same coefficient functions in front
of products of $q_\infty$ as the effective Hamiltonian $H_\lambda$ has
in front of the unitarily equivalent products of $q_\lambda$.
Differentiating ${\cal H}_\lambda$ one obtains

$$ {d\over d\lambda} \,\, {\cal H}_\lambda \quad = \quad - \quad [{\cal
T}_\lambda, \, {\cal H}_\lambda] \quad , \eqno (3.3)$$

\noindent where the generator ${\cal T}_\lambda$ is related to ${\cal
U}_\lambda$ by

$$ {\cal T}_\lambda \quad = \quad {\cal U}^\dagger_\lambda \, {d\over
d\lambda} \, {\cal U}_\lambda \quad . \eqno(3.4) $$

\noindent The script letters are introduced to indicate that the
operators can be conveniently thought about as expanded into sums of
products of operators $q_\infty$.  The latter are independent of
$\lambda$ and are not differentiated in Eqs.  (3.3) and (3.4).  In other
words, Eqs.  (3.3) and (3.4) describe only the flow of coefficients in
front of the creation and annihilation operators.  Effective
Hamiltonians are obtained from ${\cal H}_\lambda$ using $
H_\lambda(q_\lambda) = {\cal U}_\lambda {\cal H}_\lambda {\cal
U}^\dagger_\lambda$.

The key element now is how one defines ${\cal T}_\lambda$.  This is the
domain of similarity for effective particles.  In its essence
\cite{GlazekWilson1} \cite{GlazekWilson2}, one studies what one has to
do to get the narrow Hamiltonian matrices as a result of the procedure,
and these studies tell us what to put for ${\cal T}_\lambda$.  There
exist infinitely many choices.  The one that the present author used to
get results described in the next Sections is of the following form
\cite{GlazekAPP} \cite{GlazekRC}

$$ [{\cal T}_\lambda, \, {\cal H}_{0\lambda}] \quad = \quad {d \over d
\lambda} \, (1 - F_\lambda)[{\cal G}_\lambda] \quad . \eqno(3.5) $$

\noindent The symbols $\cal G$ and $F$ require explanation.  The
effective Hamiltonian $H_\lambda$ contains form factors of width
$\lambda$ in all its vertices.  If we denote an operator without the
form factors by $G_\lambda$, our Hamiltonian takes the form $H_\lambda =
F_\lambda[G_\lambda]$, where the operator $F_\lambda$ inserts the form factors.
With these form factors, momentum transfers in interactions between
effective particles are guaranteed to be at most of the order of
$\lambda$.  ${\cal G}_\lambda = {\cal U}^\dagger_\lambda G_\lambda {\cal
U}_\lambda$.  We divide ${\cal G}_\lambda$ into two parts, a part that
is bilinear in $q_\infty$, and an interaction part that would vanish if
the coupling constant were equal 0, so that ${\cal G}_\lambda = {\cal
G}_0 + {\cal G}_{I\lambda}$.  The operator ${\cal G}_{I\lambda}$
satisfies the following differential equation as a consequence of Eqs.
(3.2)-(3.5),

$$ {d \over d\lambda} {\cal G}_{I\lambda} \quad = \quad \left[ f{\cal
G}_I, \, \left\{ {d \over d\lambda} (1-f){\cal G}_I \right\} _{{\cal
G}_0} \right] \quad . \eqno(3.6) $$

\noindent We dropped the subscript $\lambda$ on the right-hand side for
clarity.  $f$ denotes the similarity form factor introduced by
$F_\lambda$ and the curly bracket with the subscript ${\cal G}_0$
denotes a solution for $ {\cal T}_\lambda$ resulting from Eq.  (3.5).

\vskip.3in
{\bf 4. ASYMPTOTIC FREEDOM IN SCALAR THEORY}
\vskip.1in

Since the interaction term (1.1) is only a part of the QCD Hamiltonian
and the function $Y_{123}$ depends on spins and momenta of gluons,
let us first discuss the case of scalar field with classical Lagrangian
density

$$ {\cal L} = {1 \over 2} ( \partial_\mu \phi \partial^\mu \phi - \mu^2
\phi^2) - {g \over 3 !  } \phi^3 \quad . \eqno(4.1) $$

\noindent In this case, the interaction term in the corresponding
Hamiltonian is of the form (1.1), but $Y_{123} = 1/2$ and calculations
are much simpler than in QCD.  Our goal is to describe results for the
light-front Hamiltonian for effective bosons calculated in perturbation
theory up to third power in $g$.  Although our presentation is based on
Ref.  \cite{GlazekRC} that uses plain expansion in powers of $g$, the
reader may also wish to compare our results with Ref.
\cite{AllenPerry}, where a different scheme is used, including
transverse locality and coupling coherence.  \cite{BrisudovaPerryWilson}
\cite{AllenAdelaide}

The light-front Hamiltonian corresponding to the Lagrangian density
(4.1) reads

$$ H_\infty \quad = \quad \int [k] {k^{\perp \, 2} + \mu^2 \over k^+}
a^\dagger_{\infty k} a_{\infty k} \quad + $$
$$+ \quad {g \over 2} \int
[k_1 k_2 k_3] \, 2(2\pi)^5 \delta^5(k_1 + k_2 - k_3)\,(a^\dagger_{\infty k_1}
a^\dagger_{\infty k_2} a_{\infty k_3} + a^\dagger_{\infty k_3} a_{\infty
k_2} a_{\infty k_1} ) \, r_\Delta \,\, + \,\, X_\Delta \quad , $$
$$ \eqno(4.2) $$

\noindent where $r_\Delta$ is a smooth regularization factor and $X_\Delta$
denotes counterterms (derivable in perturbation theory). In $n$ dimensions,
$[k]$ means $\theta(k^+) dk^+ d^{n-2} k^\perp/(2k^+ (2\pi)^{n-1})$. We choose

$$ r_\Delta \quad = \quad \exp{- (\eta_1 + \eta_2) \, \kappa^{\perp \,
2}_{12} \over \Delta^2 } \quad , \eqno(4.3) $$

\noindent where $x_1=k_1^+/k_3^+$ and $\kappa_{12}^\perp = k_1^\perp -
x_1 k_3^\perp$, $\eta_i = \eta(x_i)$, and $\eta$ is a useful function of
its argument.  A natural choice is $\eta(x) = 1$, for it is simple.
Leaving $\eta$ unspecified will help us identify finite regularization
effects.

The similarity form factor for an operator containing $u$ creation operators
and $v$ annihilation operators is defined by

$$ f_\lambda(u, v) \quad = \quad \exp{ [ - ({\cal M}^2_{u} - {\cal
M}^2_{v})^2/\lambda^4]} \quad . \eqno(4.4) $$

\noindent The script notation for invariant masses means
$ {\cal M}^2_u = (k_1 + ...  + k_u)^2$, where the minus components
of the momentum four-vectors are given by $k^-_i = (k^{\perp\,2}_i +
\mu^2)/k^+_i$ for $i = 1, ..., u$, and similarly for $v$.

Equation (3.6) can now be solved order by order using expansion in
powers of $g$.  Firstly, one obtains the counterterms $X_\Delta$ as the
initial conditions at $\lambda = \infty$ that render regularization
independent finite $\lambda$ Hamiltonians.  To order $g^3$, the
regularization dependence of $H_\lambda$ lets us identify two
counterterms:  the mass counterterm

$$ \beta_{\infty 11} = \int [k] \, {\delta \mu^2_\infty \over k^+}
a^\dagger_{\infty k} a_{\infty k} \quad , \eqno(4.5) $$

\noindent and the vertex counterterm

$$ \gamma_{\infty 2 1} = \int [k_1 k_2 k_3] \, 2(2\pi)^5 \delta^5 (k_1
+ k_2 - k_3) \, \gamma_\infty (k_1, k_2, k_3)
\, a^\dagger_{\infty k_1} a^\dagger_{\infty k_2}
a_{\infty k_3} \, r_\Delta \quad . \eqno(4.6) $$

Without loss of generality, we assume that some gedanken experimental data
require the mass squared parameter in effective Hamiltonian with
$\lambda = \lambda_0$ to be equal $\mu^2 + \delta \mu^2_0$. This means
that when one calculates observables using the effective Hamiltonian,
$\mu^2_{\lambda_0}$ must equal $\mu^2 + \delta \mu^2_0$ to fit the data.
This condition, by tracing the renormalization group equation for
$H_\lambda$ back to $\lambda = \infty$, tells us that

$$ \delta \mu^2_\infty = \delta \mu^2_0 - \left({g \over 2}\right)^2
{1\over 2 (2\pi)^5} \int_0^1 {dx \over x(1-x) } \int d^4 \kappa^\perp {2
\over {\cal M}^2 - \mu^2} \left[ f^2_{\lambda_0} ({\cal M}^2, \mu^2) -
1\right] \, r_{\Delta \beta}\quad . \eqno(4.7) $$

\noindent The script ${\cal M}$ denotes invariant mass, ${\cal M}^2 =
(\kappa^{\perp \, 2} + \mu^2)/x(1-x)$, and the regularization factor is

$$ r_{\Delta \beta} \quad = \quad \exp{ \left\{-2[\eta(x) +
\eta(1-x)]\kappa^{\perp \,2}/\Delta^2 \right\} } \quad . \eqno(4.8) $$

\noindent Integration gives two diverging terms, one proportional to
$\Delta^2$ and another one proportional to $\log{\Delta}$.  The
remaining finite part depends on our choice of the function $\eta$.
For example, evaluating the integral for $\eta(x) = 1/x$ one obtains

$$ \delta \mu^2_\infty = g^2 {1 \over (4\pi)^3} \left[ \, {1\over
24}\Delta^2 \, - \, \mu^2 {5\over 6} \log{\Delta \over \mu} \, + \,
\mu_\eta^2 \right] \quad , \eqno(4.9) $$

\noindent where $\mu_\eta$ has a finite limit when $\Delta \rightarrow
\infty$.  The logarithmically divergent part is independent of the
function $\eta$ and agrees with results for the Lagrangian mass squared
counterterm obtained using Feynman diagrams and dimensional
regularization \cite{MacfarlaneWoo} \cite{Collins} in the following
sense:  when one changes $\Delta$ to $\Delta'$ the logarithmic part of
the counterterm changes with $\Delta$ as the mass squared changes as a
function of the renormalization scale in Eq.  (7.1.22) in
\cite{Collins}.

The vertex counterterm is defined by the requirement that the effective
vertex in the Hamiltonian $H_{\lambda_0}$ is free from regularization
dependence for arbitrary finite values of $\lambda_0$. The one loop
regularization sensitive contributions to the effective vertex function
are given by

\begin{eqnarray*}
&& \gamma_\infty(k_1, k_2, k_3) \quad + \quad
\left({g \over 2}\right)^3 {\pi^2 \over 2(2\pi)^5} \times \\
&& \\
&& \times \left[ {1\over 2} \left[ \int_{x_1}^1 {dx \over x(1-x)(x-x_1)}
\int_0^\infty \kappa^2 d\kappa^2 \,\, 8 {x-x_1 \over x x_2 {\cal M}^4 }
\exp{ \left({- c_\eta \kappa^2 \over \Delta^2}\right)} + ( x_1
\leftrightarrow x_2 ) \right] + \right. \\
&& \\
&& \left. + \int_0^1 {dx \over x(1-x)} \int_0^\infty \kappa^2 d\kappa^2
{-3 \over {\cal M}^4} \exp{ \left({- d_\eta \kappa^2
 \over \Delta^2}\right)}\right] \quad ,
\end{eqnarray*}
$$\eqno(4.10)$$

\noindent where

$$ c_\eta = \eta (x) + \eta (1-x) + \left\{ \eta (x_1/x) + \eta
[(x-x_1)/x] \right\} (x_1/x)^2 + \eta[(x-x_1)/x_2] + \eta[(1-x)/x_2] \,
 \eqno(4.11) $$

\noindent and

$$ d_\eta \quad = \quad 2 \, [\eta(x) + \eta(1-x)] \quad . \eqno(4.12)$$

\noindent The counterterm function $\gamma_\infty(k_1, k_2, k_3)$ must
remove the regularization dependence from the above expression.  The
regularization effects are independent of $\kappa^\perp_{12}$.  Dropping
all parts that are independent of regularization, we conclude that

\begin{eqnarray*}
&&  \gamma_\infty (k_1, k_2, k_3) \quad + \quad \left({g \over 2}\right)^3
{1 \over (4\pi)^3} \times \\
&& \\
&& \left[ 3 \log{\Delta \over \mu} \,
- \, 4 \left[ \int_{x_1}^1 dx {1-x \over x_2} \log{c_\eta}
\, + \, (x_1 \leftrightarrow x_2) \right]
\, + \, 3 \int_0^1 dx \, x(1-x)
\log{d_\eta} \, \right] \quad .
\end{eqnarray*}
$$\eqno(4.12) $$

\noindent must be independent of regularization.  We see that the
diverging regularization dependence of the interaction vertex, i.e. the
term proportional to $\log {\Delta}$, is independent of the particle
momenta and one can remove the divergence by introducing a
$\gamma_\infty (k_1, k_2, k_3)$ that is equivalent to changing the
initial coupling constant $g$ in Eq.  (4.2).  Thus, no diverging
$x$-dependent counterterms are required - a different situation than in
\cite{overlap}.  However, it is also visible that the vertex contains a
finite regularization dependent part that is a function of $x_1$.  The
function depends on our choice for $\eta$.  For example, if $\eta = 1$
one has $c_\eta = 4 + 2(x_1/x)^2$ and $d_\eta = 4$.  The resulting
integral is a function of $x_1$, and needs to be subtracted.  But this
would not assure us that the whole ultraviolet regularization dependence
is removed, because we work with a specific functional form of the
regulating function (4.3).

Since the whole regularization effect is independent of $\lambda$ and
$\kappa^\perp_{12}$, it can be completely removed from the effective
interaction by subtracting its value for $\kappa^\perp_{12} = 0$ at an
arbitrarily chosen finite $\lambda_0$.  However, one has to add back the
finite regularization independent part of the effective vertex, which is
a function of $x_1$, denoted below by $\gamma_0(x_1)$.  The function
$\gamma_0(x_1)$ is necessary to recover Poincar\'e symmetry of
observables, because our regularization spoiled the symmetry.  The symmetry may
be restored once counterterms remove the regularization effects, but one
is not allowed to change terms independent of the regularization, which
were given by the initial covariant Lagrangian density unambiguously.  Therefore, the function $\gamma_0(x_1)$ must be reinserted.  This function is not altered when $\lambda$ changes and could be considered marginal in
analogy with usual renormalization group analysis.  The ultimate
adjustment of the function $\gamma_0(x_1)$ requires 4th order
calculations.  For there exists in $\phi^3$ theory no 3rd order
scattering amplitude one could use to find out what function
$\gamma_0(x_1)$ renders Poincar\'e symmetry of scattering observables
with our choice of $r_\Delta$ in Eq.  (4.2).  However, it should be
pointed out that the function does not influence the way the 3rd order
running coupling constant in effective Hamiltonians depends on
$\lambda$.

Thus, in Eq.  (4.6), the counterterm function $ \gamma_\infty (k_1, k_2,
k_3) \equiv \gamma_\infty (x_1, \kappa_{12}^\perp)$, which removes the
regularization dependence from the effective vertex reads

$$ \gamma_\infty (x_1, \kappa_{12}^\perp) \quad =
\quad - \,\, \gamma_{\lambda_0}(x_1,
0^\perp ) \,\, + \,\, \gamma_0(x_1) \quad . \eqno(4.13) $$

\noindent This result is used to define the new regularization dependent
coupling constant $g_\Delta$ in the initial Hamiltonian in Eq.  (4.2).
We select a convenient value of $x_1 = x_0$ and obtain

$$ {g_\Delta \over 2} \quad = \quad {g \over 2} \, + \,
\gamma_\infty(x_0, 0^\perp) \quad = \quad {g \over 2} \, - \,
\gamma_{\lambda_0}(x_0, 0^\perp) + \gamma_0 \quad , \eqno(4.14) $$

\noindent where $\gamma_0 \equiv \gamma_0(x_0)$.  We
see that the initial coupling $g$ is replaced by a new
$\Delta$-dependent quantity

$$ g_\Delta = g \left[ 1 - g^2 {3 \over 4(4\pi)^3 } \log{\Delta \over
m_0} \right] + o(g^5) \quad , \eqno(4.15) $$

\noindent with certain free constant $m_0$.  Thus, the theory exhibits
asymptotic freedom in 3rd order terms.  Our result agrees with
literature, say Eq.  (7.1.26) from \cite{Collins}, in the sense that
when we change $\Delta$, the change required in the coupling constant in
the initial Hamiltonian for obtaining $\Delta$-independent effective
Hamiltonians matches the change implied by Feynman diagrams and
dimensional regularization.

Having established the structure of counterterms we can proceed to
evaluation of the finite similarity flow of effective Hamiltonians
towards small widths $\lambda$. The effective kinetic energy term
in narrow Hamiltonians is

$$ H_{\lambda 11} = \int [k] \, { k^{\perp \, 2}
+ \mu^2_\lambda \over k^+}
a^\dagger_{\lambda k} a_{\lambda k} \quad , \eqno(4.16) $$

\noindent where

\begin{eqnarray*}
&&  \mu_\lambda^2 \quad = \quad \mu^2 \, + \, \delta \mu^2_\lambda \quad
= \\
&& \\
&& = \mu^2 + \delta \mu^2_0 + \left({g \over 2}\right)^2 {1\over 2
(2\pi)^5} \int_0^1 {dx \over x(1-x) } \int d^4 \kappa^\perp {2 \over
{\cal M}^2 - \mu^2} \left[ f^2_\lambda ({\cal M}^2, \mu^2) -
f^2_{\lambda_0} ({\cal M}^2, \mu^2) \right] \quad .
\end{eqnarray*}
$$\eqno(4.17) $$

\noindent The above result is particularly simple for $\mu = 0$ and in
that case it reads ($\delta \mu^2_0$ is proportional to $g^2$)

$$ \mu_\lambda^2 \quad = \quad \delta \mu^2_0 \, + \, g^2 {1\over
(4\pi)^3} \, {1\over 24} \, \sqrt{\pi \over 2} \, (\lambda^2 -
\lambda_0^2) \quad . \eqno(4.18) $$

\noindent Logarithmic dependence on $\lambda$ arises for $\mu > 0$.  The
value of $\delta \mu^2_0$ could be found, for example, by solving a
single physical boson eigenvalue problem, expressing the physical boson
mass in terms of $\delta \mu^2_0$ and adjusting the latter to obtain the
gedanken experimental mass value for bosons.  Note a change in the mass
function of cutoff parameter, from the case of the mass counterterm,
dependent on $\Delta$, to the case of running mass term, dependent on
the width $\lambda$ (independent of $\Delta$).  The change corresponds
to a transition from the initial side of a fixed point (bare canonical
Hamiltonian with regularization) to the other side of the fixed point
(renormalization group trajectory of effective Hamiltonians in the
similarity procedure, cf.  \cite{Wilson2}).

The effective vertex reads

$$ H_{\lambda 2 1} = \int [k_1 k_2 k_3] \, 2(2\pi)^5 \delta^5 (k_1 + k_2
- k_3) \,\, f_\lambda[(k_1 + k_2)^2, k_3^2] \,\,
V_\lambda(x_1,\kappa^\perp_{12} ) \,\, a^\dagger_{\lambda k_1}
a^\dagger_{\lambda k_2} a_{\lambda k_3} \quad , \eqno(4.19) $$

\noindent where $ V_\lambda(x_1, \kappa^\perp_{12})$ is the effective
vertex function and $f_\lambda$ is the similarity vertex form factor.
The vertex function is given by an integral over loop variables $x$ and
$\kappa^\perp$ of a known function.  \cite{GlazekRC}

We define the running coupling constant as the value of $2V_\lambda(x_1,
\kappa^\perp_{12})$ at a chosen configuration of momentum variables,
denoted by $(x_{10}, \kappa^\perp_{120})$.  In other words, $g_\lambda =
2V_\lambda (x_{10}, \kappa^\perp_{120})$.  A possible choice for
massless bosons is $x_{10} = 0$ and $\kappa^\perp_{120}=0$.  This is a
natural definition, analogous to the standard Thomson limit in the case
of electron charge in QED.  This choice greatly simplifies the
integrand, giving $ V_\lambda(0, 0^\perp)$, so that the result can be
fully produced here ($g_0$ is the value of $g_{\lambda_0}$ required by
phenomenology done using $H_{\lambda_0}$)

\begin{eqnarray*}
&& g_\lambda \quad = \quad g_0 \quad + \quad g_0^3 {1\over 24} {1 \over (4\pi)^3}\int_0^\infty {dz\over z}\\
&& \\
&& \left[ 2(f_\lambda - f_\lambda^3) - 2(f_0 - f_0^3) +
20(f_\lambda^3 - f_\lambda^2) - 20(f_0^3 - f_0^2) + 9(f_0^2 -
f_\lambda^2)\right] \quad ,
\end{eqnarray*}
$$\eqno(4.20) $$

\noindent where $f_\lambda = \exp{ - z^2/\lambda^4}$ and $f_0 = \exp{ -
z^2/\lambda_0^4}$.  A straightforward integration gives

$$ g_\lambda \quad = \quad g_0 \,\, - \,\, g_0^3 \,\, { 3 \over 4(4\pi)^3}
\, \log{\lambda \over \lambda_0} \quad , \eqno(4.21) $$

\noindent which exhibits asymptotic freedom.  Differentiating with
respect to $\lambda$ and keeping terms up to order $g_\lambda^3$ one
obtains

$$ {d \over d \lambda} \,g_\lambda \quad = \quad - \quad g^3_\lambda
\,\, {3 \over 256 \pi^3} \,\, {1\over \lambda} \quad . \eqno(4.22) $$

\noindent This equation demonstrates the same $\beta$ function for
coupling constants in effective Hamiltonians as obtained in Lagrangian
approaches using Feynman diagrams and dimensional regularization, when
one identifies the renormalization scale with the Hamiltonian width
$\lambda$.  This is encouraging but one needs to remember that for
comparison of perturbative scattering amplitudes in Hamiltonian and
Lagrangian approaches it is necessary to make additional calculations
and at least of fourth order in $g$.  Beyond model matrix studies such
as in \cite{GlazekWilson3}, 4th order similarity calculations have so
far been carried out only in a simplified Yukawa model by Mas{\l}owski and
Wi\c eckowski \cite{MaslowskiWieckowski} (the latter model calculations
should be helpful in setting up a light-front theory of nucleons and pions).

Integrating Eq.  (4.10), one obtains ($\alpha = g^2/4\pi$)

$$ \alpha_\lambda = { \alpha_0 \over 1 + \alpha_0 (3/32\pi^2)
\log{\lambda / \lambda_0} } \quad , \eqno(4.23) $$

\noindent which shows our result for a boost invariant running coupling
constant in effective Hamiltonians.  Our procedure explains how the
running coupling constant can be included in quantum mechanics of
effective particles, which is given by the Schr\"odinger equation with
the corresponding Hamiltonian $H_\lambda$.  One can evaluate matrix
elements of the small width Hamiltonian in a limited subspace of Fock
states built of the effective particles.  The form factor in the
interaction vertex (4.19) secures a small range of the interactions on
the energy scale and one can expect a rapid convergence of wave
functions in the effective particle basis.

Note three characteristic features of the Hamiltonian calculation.  (1)
No field renormalization constant appeared, since the similarity
transformation did not eliminate (or integrated out) any degrees of
freedom.  (2) No vacuum effect played any role, since we used the
light-front form of dynamics.  Extensive literature concerning the
vacuum issue can be traced through reference \cite{Susskind}.  $\phi^3$
theory is unstable due to a possibility that the field $\phi$ takes an
infinitely large, negative value.  It would be interesting to check if
the perturbatively evaluated effective Hamiltonians of small widths have
any tendency to develop eigenstates that deviate in that direction.  (3)
The Hamiltonian structure is invariant with respect to boosts, including
boosts from the rest frame of any bound state to the infinite momentum
frame.  This suggests the approach outlined above should be tried in
QCD, and in effective theories of strong interactions in nuclear
physics, to connect low energy observables, such as binding energies,
radii or magnetic moments of bound states, with high energy ones, such
as parton distributions, form factors or jets.

\vskip.3in
{\bf 5. QCD GLUON VERTEX COUNTERTERM}
\vskip.1in

We come back to Eq.  (1.1) in QCD and repeat the same analysis as we did
for the scalar theory.  Most of the procedure remains the same, but an
important complication arises.  The vertex function in the canonical
Hamiltonian has now the form ($c$ refers to color and $\epsilon$ to
polarization of gluons)

$$ Y_{123} = i f^{c_1 c_2 c_3} [ \epsilon^{*\perp}_1 \epsilon^{*\perp}_2
\cdot \epsilon^{\perp}_3 \kappa^{\perp}_{12} - \epsilon^{*\perp}_1
\epsilon^{\perp}_3 \cdot \epsilon^{*\perp}_2 \kappa^{\perp}_{12}/x_2 -
\epsilon^{*\perp}_2 \epsilon^{\perp}_3 \cdot \epsilon^{*\perp}_1
\kappa^{\perp}_{12}/x_1]\,, \eqno(5.1) $$

\noindent in which the characteristic factors of $\kappa^\perp/x$ tend
to infinity when $ x \rightarrow 0$.  Ultraviolet coupling constant
divergences in $\phi^3$ theory in 6 dimensions resulted from transverse
momentum integration $\int d^4 \kappa^\perp /\kappa^4$, where
$1/\kappa^4$ came from the two denominators of third order perturbation
theory.  In QCD in 4 dimensions, we have instead $\int d^2 \kappa^\perp
(\kappa^\perp/x)^2 /\kappa^4$.  Therefore, in QCD (more generally, in
gauge theories), we have to introduce a separate regularization of small
$x$ behavior of interaction vertices in the Hamiltonians.  For example,
in the QCD counterpart of Eq.  (4.2), we have to insert a factor,
denoted by $r_\delta$, in addition to $r_\Delta$, that will effectively
cut-off the region where one of the gluons 1 or 2 carries a smaller
fraction of $k_3^+$ than the size of a small parameter $\delta$.

The small $x$ regularization function $r_\delta$ may appear to be only a
technical detail.  But it was pointed out by Perry that singularities at
small $k^+$ may be related to effective confining potentials in
$H_\lambda$, calculable already in second order perturbation theory.
\cite{Perryconfinement} In short, the canonical light-front QCD
Hamiltonian contains terms that are singular at small $k^+$ and the
singularity contributes to the effective Hamiltonians $H_\lambda$,
providing potentials that grow with distance between color charges.
This is quite different a situation from other formulations of the
theory, where second order calculations are not expected to tell us
anything about confinement.  Therefore, the small $x$ features of QCD in
the light-front Hamiltonian approach deserve extensive studies.  Here,
we merely report some initial results for third order gluon vertex
counterterm, indicating $x$-dependent features.

The whole analysis of the previous Section can be repeated step by step
and one can derive the interaction term for effective gluons, in the
narrow Hamiltonian $H_\lambda$ for QCD.  The condition that the
effective vertex is independent of regularization parameter $\Delta$
gives us the diverging triple-gluon vertex counterterm in the initial
QCD Hamiltonian.  The new features appear in the coefficient of $\log
\Delta$.  Namely, the diverging part of the vertex, to be compensated by
the ultraviolet counterterm, has the form

$$ {g^3 \over 4\pi^2}\, \left[ {11\over 12} N_c - {1\over 6} n_f + N_c
f(x_1,x_2)\right] \, \log{\Delta} \cdot Y_{123} \quad ,\eqno(5.2)$$

\noindent where the function $f(x_1,x_2)$ is symmetric in its arguments.
This function originates from three successive actions of the triple
gluon interaction, the gluon mass correction providing only a constant
contribution, and it depends on the regularization factor $r_\delta$ in
the initial Hamiltonian.  In fact, $f(x_1, x_2) = [- \log{x_1} +
\int_0^1 dx (2/x + 1/(1-x))(r_{\delta 3} - r_{\delta 2}) + (1
\leftrightarrow 2)]/2$, where $r_{\delta 3} = r_{\delta 3}(x,x_1)$ is a
product of three factors $r_\delta$ for the three successive
triple-gluon interactions, and $r_{\delta 2} = r^2_\delta(x) $, is a
product of two factors $r_\delta$ from the gluon mass counterterm.  For
$r_\delta(x) = \theta(x-\delta)\theta(1-x-\delta)$ in the bare vertex,
for gluons carrying $x$ and $1-x$ of the single gluon momentum $k_3^+$
in Eq.  (1.1), $f(x_1,x_2) = \log[min(x_1,x_2)]$, which is negative and
hence reduces the rate at which the initial coupling depends on
$\Delta$.  The latter feature comes about as follows.  Since the
counterterm must contain the term opposite in sign to (5.2), the
coupling constant in the regularized $H_\infty(q_\infty)$ in QCD is
changed to [cf.  Eq.  (4.15)]

$$ g_\Delta \quad = \quad g \quad - \quad {g^3 \over 4\pi^2}\,
\left[ {11\over 12} N_c - {1\over 6} n_f
+ N_c f(x_0,1-x_0)\right] \, \log{\Delta} \quad . \eqno(5.3) $$

\noindent The coefficient of $\log \Delta$ is independent of the
parameter $\delta$, but it depends on the small $x$ regularization in a
finite way.  Having derived the function $f(x_1,x_2)$, for certain
choices of $x_1 = x_0$ in the definition of the Hamiltonian coupling
constant, one could obtain triviality instead of asymptotic freedom.
This is a peculiar result for the regularization given in Eq.  (4.3) and
the sharp cutoff on $x$ at $\delta$.  In the limit when both $r_{\delta
3}$ and $r_{\delta 2}$ are replaced by 1, one would obtain $f(x_1, x_2)
= - \log{\sqrt {x_1 x_2}}$, which is always positive, and would
accelerate the asymptotic freedom rate of change of $g_\Delta$ with
$\Delta$.  One can seek choices of regulating functions $r_\Delta$ and
$r_\delta$ that eliminate the unusual logarithm (the other terms are
standard) but it is not known if a finite function of $x_1$ is not
necessary in place of $f(x_1, x_2)$ in light-front Hamiltonians anyway,
to restore symmetries for physical quantities.  In addition, as in the
scalar theory, the ultraviolet finite part of the counterterm involves
an unknown function of $x_1$.  One could say that the structure of the
model from Ref.  \cite{overlap} is closer to QCD than to scalar theory
in 6 dimensions.  Evaluation of the effective coupling constant
$g_\lambda$ in QCD, in analogy to Eq.  (4.20) in scalar theory, may shed
some light on how to disentangle genuine ultraviolet from small $x$
singularities in Hamiltonians.

It is clear from the above example that effective light-front
Hamiltonians of QCD require careful studies employing various types of
regulators before we will know the optimal ways of calculating window
Hamiltonians. The interplay of transverse and longitudinal momentum variables may lead to surprising results. However, the calculations are certainly doable and the resulting matrices will tell us about details of QCD dynamics in the
Fock space of effective quarks and gluons. The similarity renormalization
group procedure for Hamiltonians is able to reveal new features of effective particle dynamics which standard Lagrangian approaches do not reveal.

\vskip.3in
{\bf 6. TRANSITION TO NEW DEGREES OF FREEDOM}
\vskip.1in

We may hope to make a transition from the effective QCD degrees of
freedom to nuclear physics hadronic interactions, such as pion-nucleon
coupling, after we achieve understanding of the narrow $H_\lambda$ in
QCD. To see the basis for this hope, let us come back again to the
matrix picture.

Once we have the narrow Hamiltonian matrix, we can divide it into the
boxes as it is illustrated in the figure below, neglecting the small
triangles in the band outside of them, which are initially left out.  We
can find eigenstates of the boxes (they correspond to different
invariant mass states) and calculate the band diagonal matrix matrix
elements in the basis built from those eigenstates.  The states
corresponding to the middle energy scale of each box will not interact
very strongly with neighboring (in energy, or mass) states, but matrix
elements sensitive to the left-out triangles will lead to strong
interactions.  We can imagine that the lowest box corresponds to
nucleons interacting through potential forces, the next box corresponds
to nucleons plus one meson, the second box to nucleons and two mesons,
etc.  This is how one could make a connection between the QCD dynamics
and nuclear physics through similarity renormalization group for
Hamiltonians (cf.  \cite{Bylev}).

\vskip.2in
\centerline{\bf Changing degrees of freedom}
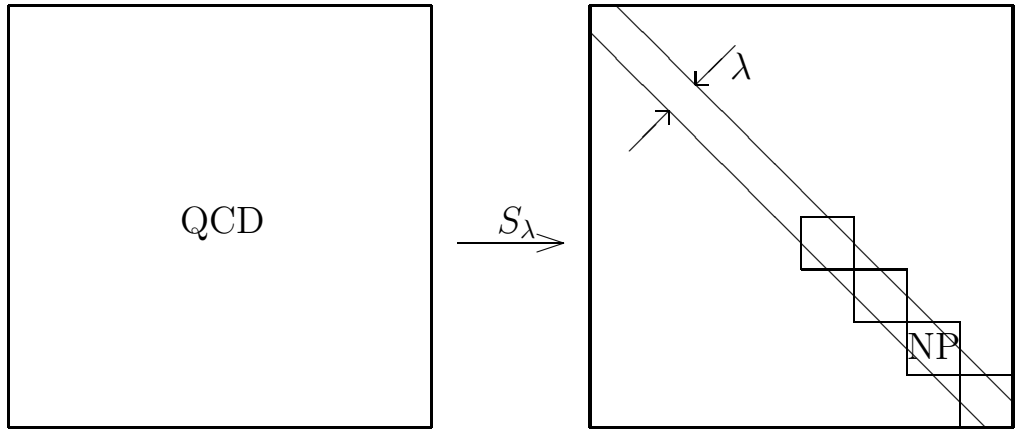
\begin{figure}[h]
\begin{center}
{\large
\begin{picture}(420,170)(-10,0)


\thicklines
\put(-10,0){\line(1,0){160}}
\put(-10,0){\line(0,1){160}}
\put(150,160){\line(-1,0){160}}
\put(150,160){\line(0,-1){160}}

\put(210,0){\line(1,0){160}}
\put(210,0){\line(0,1){160}}
\put(370,160){\line(-1,0){160}}
\put(370,160){\line(0,-1){160}}

\thinlines
\put(360,0){\line(-1,1){150}}
\put(370,10){\line(-1,1){150}}
\put(240,120){\line(-1,-1){15}}
\put(250,130){\line(1,1){15}}
\put(240,120){\line(-1,0){5}}
\put(240,120){\line(0,-1){5}}
\put(250,130){\line(1,0){5}}
\put(250,130){\line(0,1){5}}
\put(263,132){$\lambda$}

\put(160,70){\line(1,0){40}}
\put(200,70){\line(-3,1){10}}
\put(200,70){\line(-3,-1){10}}
\put(175,74){$S_\lambda$}

\put(55,74){QCD}
\put(330,26){NP}

\put(290,60){\line(1,0){20}}
\put(290,60){\line(0,1){20}}
\put(310,80){\line(-1,0){20}}
\put(310,80){\line(0,-1){20}}

\put(310,40){\line(1,0){20}}
\put(310,40){\line(0,1){20}}
\put(330,60){\line(-1,0){20}}
\put(330,60){\line(0,-1){20}}

\put(330,20){\line(1,0){20}}
\put(330,20){\line(0,1){20}}
\put(350,40){\line(-1,0){20}}
\put(350,40){\line(0,-1){20}}

\put(350,0){\line(1,0){20}}
\put(350,0){\line(0,1){20}}
\put(370,20){\line(-1,0){20}}
\put(370,20){\line(0,-1){20}}


\end{picture} } \end{center} \end{figure} However, this is not the only
possibility one can try to explore for the change of basis.  One can
consider new basis states built of quarks and gluons, possibly open
gluon string bits with quarks at the ends or closed rings of gluons, and
evaluate matrix elements of the effective Hamiltonians between such
objects.  Since one has a perturbative expression for the operators
${\cal U}_\lambda$, see Eqs.  (3.1) and (3.2), one can attempt
evaluation of matrix elements between states that are constructed in a
variety of ways, using quarks and gluons corresponding to different
scales $\lambda$.  One could even ask if there is a way to calculate a
connection between the quark and gluon matrices of intermediate widths
and reggeized gluon interactions, once one restricts the space of states
to those that dominate in multi-Regge kinematics.  \cite{Lipatov}

\vskip.3in
{\bf 7. CONCLUSION}
\vskip.1in

Asymptotically free theories can be analyzed using Hamiltonian approach.
The analysis can be based on the similarity renormalization group
procedure for effective particles.  The evaluation of running couplings
in the effective Hamiltonians can be carried out without introduction of
wave function renormalization constants and without invoking any
properties of the vacuum state (in the light-front form of Hamiltonian
dynamics).  In third order calculations, one obtains familiar asymptotic
results in scalar $\phi^3$ theory, plus an $x$-dependent finite
counterterm.  In QCD, the standard asymptotic freedom form of
triple-gluon vertex counterterm is supplemented by an ultraviolet
diverging and $x$-dependent counterterm, and by an ultraviolet finite
$x$-dependent counterterm.  Effects predicted for QCD by the power
counting in $k^\perp$ and $k^+$ \cite{Wilsonetal} are confirmed but the
analysis is changed by transition to boost invariant variables
$\kappa^\perp$ and $x$, and detailed calculations may produce results
that are not expected to emerge from Feynman diagrams.  Mixing between
the small $x$ and large $\kappa^\perp$ cutoffs indicates a need for a
new precise definition of the ultraviolet domain in the Hamiltonian
approach.  Nevertheless, one obtains well-defined expressions for
effective Hamiltonian interactions without necessity to calculate
scattering matrix elements for quarks and gluons as if they were
observable particles.

The effective particle calculus preserves cluster properties and allows
for evaluation of effective Hamiltonians without limitation to any
particular set of matrix elements.  In other words, we can derive
integral expressions for matrix elements of effective Hamiltonians in
the whole Fock space spanned by basis states of effective particles.
The renormalization group equations are integrated analytically using
gaussian similarity form factors and one fully controls off-shell
behavior of effective vertices that correspond to the initial theory.
The effective dynamics is invariant with respect to boosts and allows
simultaneous analysis of the rest frame and infinite momentum frame
structure of bound states.

The effective particle Fock space expansion can converge thanks to the
similarity form factors in the interaction vertices.  The form factors
dampen interactions changing invariant masses by more than $\lambda$ and
thus can tame the spread of eigenstate wave functions for low lying
eigenvalues into regions of high relative momenta of constituents.  This
feature may lead to exponential convergence of the eigenstate expansion
in the effective particle basis.  Such convergence is not expected in
the case of bare particles.  The fine structure of effective particles
would then unfold in the transformation ${\cal U}_{\lambda_1} {\cal
U}^\dagger_{\lambda_2}$ relating effective degrees of freedom at two
different scales, one corresponding to the binding scale and the other
to the high momentum transfer probe in question.

The near-diagonal Hamiltonian matrices allow for transition to new
degrees of freedom by turning to basis states that are eigenstates of
small block matrices on the diagonal.  In principle, these new degrees
of freedom could correspond to mesons and baryons built from constituent
quarks and gluons, in case of QCD.  One can also consider other changes
of basis states and seek most efficient degrees of freedom, such as
strings of gluons with quarks at the ends, for solving the
nonperturbative eigenvalue problems for narrow effective Hamiltonians.

\vskip.3in
{\bf Acknowledgments}
\vskip.2in

The author thanks Lev Lipatov for discussions about $x$-dependent effects and Ken Wilson for comments and discussion about the manuscript.  Support and hospitality from Andreas Schreiber, Tony Thomas, Tony Williams, and Frederic Bonnet during the author's stay at CSSM, are gratefully acknowledged.

\end{document}